\documentclass[journal,transmag]{IEEEtran}

\usepackage{amsmath}
\usepackage{graphicx}

\begin{document}

\title{Non-collinear Magnetic Configurations at Finite Temperature in Thin Films}

\author{\IEEEauthorblockN{Levente R\'ozsa\IEEEauthorrefmark{1},
L\'aszl\'o Szunyogh\IEEEauthorrefmark{1}\IEEEauthorrefmark{2},
and L\'aszl\'o Udvardi\IEEEauthorrefmark{1}\IEEEauthorrefmark{2}}
\IEEEauthorblockA{\IEEEauthorrefmark{1}Department of Theoretical Physics, Budapest University of Technology and Economics,\\~H-1111 Budapest, Hungary}
\IEEEauthorblockA{\IEEEauthorrefmark{2}Condensed Matter Research Group of Hungarian Academy of Sciences,\\
Budapest University of Technology and Economics,
~H-1111 Budapest, Hungary}
\thanks{Manuscript received March 6, 2014. Corresponding author: L. R\'ozsa (email: rozsa@phy.bme.hu).}}

\markboth{Intermag 2014 session BE-04 Digest ID 1361}%
{Shell \MakeLowercase{L. R\'ozsa \textit{et al.}}: Non-collinear magnetic configurations at finite temperature in thin films}

\IEEEtitleabstractindextext{%
\begin{abstract}
The finite-temperature magnetism of a monolayer on a bcc (110) surface was examined using a model Hamiltonian containing ferromagnetic or antiferromagnetic exchange interactions, Dzyaloshinsky-Moriya interactions and easy-axis on-site anisotropy. We examined the competition between the collinear ground state parallel to the easy axis and the spin spiral state in the plane perpendicular to this axis preferred by the Dzyaloshinsky-Moriya interaction. Using approximative methods to calculate the magnon spectrum at finite temperatures, it was found that even if the ground state is collinear, increasing the Dzyaloshinsky-Moriya interaction strongly decreases the critical temperature where this collinear order disappears. Using atomistic spin dynamics simulations it was found that at this critical temperature the system transforms into the non-collinear state. Including external magnetic field helps stabilising the ferromagnetic state. An effect due to the finite size of the magnetic monolayer was included in the model by considering a different value for the anisotropy at the edges of the monolayer. This effect was shown to stabilize the spin spiral state by fixing the phase at the ends of the stripe.
\end{abstract}

}

\maketitle

\IEEEdisplaynontitleabstractindextext

\IEEEpeerreviewmaketitle

\section{Introduction}

\IEEEPARstart{R}{ecent} experiments on magnetic thin films have provided evidence for the presence of interesting non-collinear magnetic configurations. A Mn monolayer deposited on W(110) surface has a homogeneous spin spiral ground state as shown by spin-polarized scanning tunnelling microscopy (SP-STM) experiments\cite{Bode}. SP-STM experiments also helped in determining that double-layer Fe on W(110) shows spiral ordering, though a largely inhomogeneous one resembling a sequence of domain walls\cite{Meckler}. It turns out that Dzyaloshinsky-Moriya (DM) interactions\cite{Dzyaloshinsky}\cite{Moriya} as a consequence of the absence of inversion symmetry of the surface and the strong spin-orbit coupling in the substrate have an important role in determining the rotational sense and the wavelength of the spiral state. Moreover, in the case of a ferromagnetic ground state as in Fe monolayer on W(110), the DM interaction leads to an asymmetry in the magnon spectrum around the $\Gamma$ point\cite{Udvardi}, with experiments\cite{Zakeri}\cite{Zakeri2} also confirming the presence of this asymmetry.

Another effect of the DM interaction is that it can induce a transition between commensurate and incommensurate states as a function of temperature. Although these effects are widely discussed in the literature in the case of three-dimensional systems\cite{Dzyaloshinsky2}\cite{Izyumov}\cite{Kousaka}, the finite-temperature behaviour of two-dimensional films is less explored from this viewpoint. Therefore, in this paper we examine finite-temperature effects on a magnetic monolayer with the symmetry of a bcc (110) surface.

\section{Theoretical model}

\begin{figure}
\centering
\includegraphics[width=0.8\columnwidth]{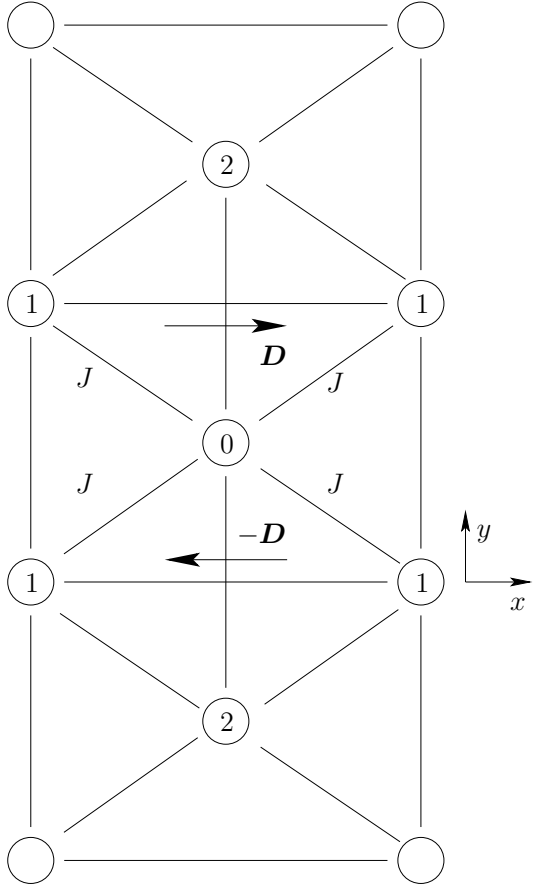}
\caption{The (110) surface of a bcc lattice. The nearest neighbours of site 0 are denoted by 1, the next-nearest neighbours by 2. The $J$ exchange parameters between 0 and 1, as well as the $\boldsymbol{D}$ DM vectors between 0 and 2, are also shown.\label{fig1}}
\end{figure}

We use an atomistic model Hamiltonian,
\begin{eqnarray}
H&=&\sum_{\langle i,j \rangle_{1}}J\boldsymbol{\sigma}_{i}\boldsymbol{\sigma}_{j}+\sum_{\langle i,j \rangle_{2}}\boldsymbol{D}_{ij}\left(\boldsymbol{\sigma}_{i}\times\boldsymbol{\sigma}_{j}\right)\nonumber
\\
&&+\sum_{i}K\sigma_{ix}^{2} - \sum_{i}\boldsymbol{B}M\boldsymbol{\sigma}_{i},\label{Ham}
\end{eqnarray}
where $\boldsymbol{\sigma}_{i}$ is a classical unit vector representing the spin at site $i$, $J$ is the isotropic exchange coupling between the nearest neighbours which can either be ferromagnetic or antiferromagnetic, $\boldsymbol{D}_{ij}$ is the DM vector between the next-nearest neighbours, while $K<0$ is an on-site anisotropy constant designating an easy $x$ axis. $\boldsymbol{B}$ denotes the external magnetic field and $M$ stands for the size of the magnetic moments in the monolayer. The $x$, $y$ and $z$ axes are parallel to the $[1\overline{1}0]$, $[001]$ and $[110]$ directions, respectively. The coordinate system, the atomic positions and the interactions are sketched in Fig.~\ref{fig1}. Since the next-nearest neighbours are located in the $y-z$ plane, which is a symmetry plane of the system, it can be shown by symmetry arguments\cite{Moriya2} that the $\boldsymbol{D}_{ij}$ vector must be parallel to the $x$ axis. Since the DM interaction favors if the spins are oriented perpendicular to the $\boldsymbol{D}_{ij}$ vector, it has a competing effect with the easy-axis anisotropy $K$. The magnetic states of the system are examined as a function of temperature, external magnetic field and the values of the parameters in the Hamiltonian (\ref{Ham}).

\section{The magnon spectrum of the system}

\begin{figure}
\centering
\includegraphics[width=\columnwidth]{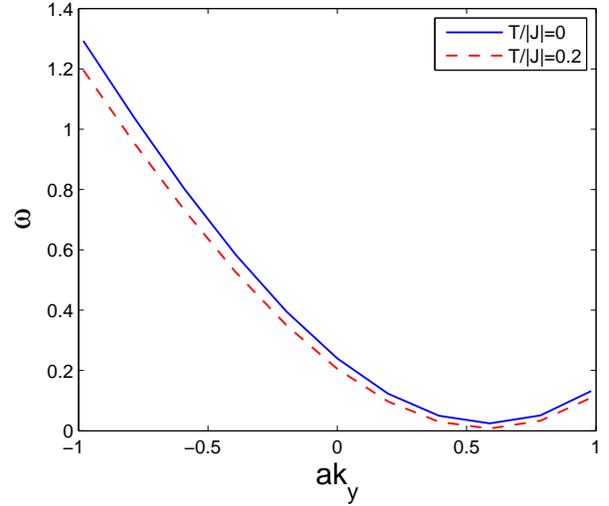}
\caption{The magnon spectrum of the model Hamiltonian (\ref{Ham}), for the parameters $J=-1$, $D=0.349$ and $K=-0.12$, at two different temperatures. Note that the minimum is not at $ak_{y}=0$ and that increasing the temperature decreases the magnon energies.\label{fig2}}
\end{figure}

By using the Landau-Lifshitz equations\cite{Landau}, which are known to give a relatively good description of the motion of the spins if this motion is considerably slower than the electronic processes\cite{Antropov}, the spectrum of the classical spin waves (magnons) can be calculated\cite{Udvardi2}. For ferromagnetic coupling, $J<0$ in (\ref{Ham}), the spectrum of the low-energy excitations around the state where all spins point towards the $x$ axis takes the form
\begin{eqnarray}
\omega(\boldsymbol{k})&=&-4J\left[1-\cos\left(\frac{\sqrt{2}}{2}ak_{x}\right)\cos\left(\frac{1}{2}ak_{y}\right)\right]\nonumber
\\
&&-2D\sin(ak_{y})-2K, \label{spekeq}
\end{eqnarray}
with $a$ being the lattice constant of the bcc lattice, corresponding to the distance between the next-nearest neighbours, and $D$ is the magnitude of the DM vector. As obvious from (\ref{spekeq}) and visible in Fig.~\ref{fig2}, the spectrum becomes asymmetric around the center of the Brillouin zone due to the presence of the DM interaction, and a gap is induced due to the anisotropy term. If all magnon energies are positive, the system stays in the ferromagnetic ground state. On the other hand, if the minimum of the spectrum reaches $\omega=0$ and the gap disappears, the ground state will be an incommensurate spin spiral state in the $y-z$ plane. The transition between the two states occurs at approximately $D\approx\sqrt{JK}$, as can be calculated in the long-wavelength limit.

The magnon spectrum for the antiferromagnetic model, $J>0$, reads
\begin{eqnarray}
\omega_{\pm}^{2}(\boldsymbol{k})&=&\left[4J-2K \pm 2D \sin\left(ak_{y}\right) \right]^{2}\nonumber
\\
&&-16J^{2}\cos^{2}\left(\frac{\sqrt{2}}{2}ak_{x}\right)\cos^{2}\left(\frac{1}{2}ak_{y}\right),
\end{eqnarray}
where the $\pm$ sign denotes the two branches of the magnon spectrum appearing due to having two atoms in the magnetic unit cell in the antiferromagnetic ground state. Again the presence of the DM interaction shifts the minimum of the spectrum away from the $\Gamma$ point, leading to complex frequencies at approximately $D \approx \sqrt{-JK}$, indicating that the antiferromagnetic state becomes unstable.

Even if the ground state is collinear, it is important to note that the gap in the $\Gamma$ point is determined by the anisotropy, while the difference between the frequencies $\omega(\boldsymbol{k}=\boldsymbol{0})$ and $\omega(\boldsymbol{k}=\boldsymbol{k}_{min})$ is determined by the ratio of the DM interaction and the exchange term. At finite temperature it is well known that the temperature-dependent anisotropy energy decreases faster than the magnetization -- see for example \cite{Asselin} --, while the temperature-dependent exchange generally decreases at the same rate as the magnetization as can be obtained in the random phase approximation method\cite{Tyablikov}. Therefore it may be expected that the system turns into the spin spiral state at finite temperature when the temperature-dependent parameters begin to favor this type of order, before reaching a paramagnetic state.

Starting from the ferromagnetic ground state ($J<0$), one method for calculating the magnon spectrum at finite temperature is given by Bloch\cite{Bloch}, which is shown to give a good approximation\cite{Rozsa} for the magnon energies at low temperature for the classical system described by (\ref{Ham}). The result is given by
\begin{eqnarray}
\omega_{\boldsymbol{k}}(T)&=&-4J\big(1-\gamma_{\boldsymbol{k}}^{(1)}\big)-2K-2D\gamma_{\boldsymbol{k}}^{(2)}\nonumber
\\
&&+4J\frac{1}{N}\sum_{\boldsymbol{k}'}\big(1+\gamma_{\boldsymbol{k}-\boldsymbol{k}'}^{(1)}-\gamma_{\boldsymbol{k}'}^{(1)}-
\gamma_{\boldsymbol{k}}^{(1)}\big)n_{\boldsymbol{k}'}(T)\nonumber
\\
&&+4K\frac{1}{N}\sum_{\boldsymbol{k}'}n_{\boldsymbol{k}'}(T)\nonumber
\\
&&+2D\frac{1}{N}\sum_{\boldsymbol{k}'}\big(\gamma_{\boldsymbol{k}'}^{(2)}+
\gamma_{\boldsymbol{k}}^{(2)}\big)n_{\boldsymbol{k}'}(T),\label{scons1}
\\
n_{\boldsymbol{k}}(T)&=&\frac{T}{\omega_{\boldsymbol{k}}(T)},\label{scons2}
\end{eqnarray}
with $n_{\boldsymbol{k}}(T)$ the occupation number of a given energy level, $\gamma_{\boldsymbol{k}}^{(1)}=\cos(\frac{\sqrt{2}}{2}k_{x}a)\cos(\frac{1}{2}k_{y}a)$, $\gamma_{\boldsymbol{k}}^{(2)}=\sin(k_{y}a)$ geometrical factors and $N$ the number of $\boldsymbol{k}$ points in the Brillouin zone. The solution of the system of equations (\ref{scons1})-(\ref{scons2}) can be found iteratively, whereas the iterations no longer converge when the critical temperature is reached. An example for the finite-temperature solution is shown in Fig.~\ref{fig2}, where the decrease of the magnon energies at $T/|J|=0.2$ is visible compared to $T/|J|=0$. The critical temperature can also be obtained by using Tyablikov's\cite{Tyablikov} or Callen's\cite{Callen} method, giving comparable values. Regardless of the chosen method, increasing the DM interaction decreases the critical temperature, while the methods give no indication about the magnetic structure above the critical temperature.

\section{Spin dynamics simulations}

\begin{figure}
\centering
\includegraphics[width=\columnwidth]{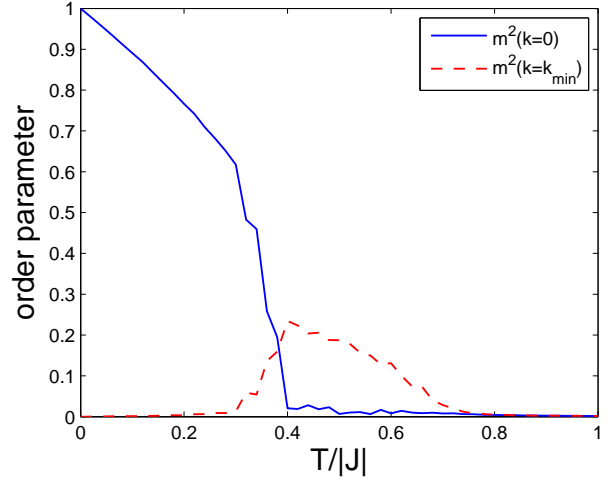}
\caption{The order parameters of the ferromagnetic ($m^{2}(\boldsymbol{k}=\boldsymbol{0})$) and the spin spiral ($m^{2}(\boldsymbol{k}=\boldsymbol{k}_{min})$) magnetic structures as a function of temperature. The system shows two consecutive phase transitions: turning from the ferromagnetic to the spiral phase at $T \approx 0.4$, and from the spiral phase to the paramagnetic phase at $T \approx 0.7$. We used the parameters $J=-1$, $D=0.349$, $K=-0.12$, and the number of atoms in the lattice was $N=64 \times 64$.\label{fig3}}
\end{figure}

In order to examine the behaviour of the system in the whole temperature range, we performed atomistic spin dynamics simulations, which are based on the solution of the stochastic Landau-Lifshitz-Gilbert equation\cite{Landau}\cite{Gilbert}\cite{Brown}\cite{Kubo}. The quantity
\begin{eqnarray}
m^{2}(\boldsymbol{k})=\langle\sum_{\alpha=x,y,z}\left|\frac{1}{N}\sum_{j}e^{i\boldsymbol{k}\boldsymbol{R}_{j}}\sigma_{j\alpha}\right|^{2}\rangle \label{orpar}
\end{eqnarray}
can be used as an order parameter to describe the magnetic structure of the system: (i) it should have a finite value at $\boldsymbol{k}=\boldsymbol{0}$ in the ferromagnetically aligned case (or at the end of the Brillouin zone in the antiferromagnetically aligned case) and disappear for all other $\boldsymbol{k}$; (ii) in the spin spiral phase it should have a finite value only for $\boldsymbol{k}=\boldsymbol{k}_{min}$; (iii) $m^{2}(\boldsymbol{k})$ should disappear in the paramagnetic phase. Exactly such a behaviour is shown in Fig.~\ref{fig3} for the ferromagnetic system, and similar results were achieved for the antiferromagnetic system. The simulations were performed on a lattice contatining $N=64 \times 64$ atoms, with periodic boundary conditions.

This type of sequence of phase transitions is slightly different from the one described in \cite{Dzyaloshinsky2} and \cite{Izyumov}, where the spins remain in the same plane in both ordered phases, but due to the anisotropy in this plane, the system switches between the commensurate and incommensurate phases. In our model, easy-axis anisotropy was only considered in the direction perpendicular to the plane of the spin spiral, parallel to the DM vector, while this is supposedly a hard axis in \cite{Dzyaloshinsky2}-\cite{Izyumov}, since it plays no role in the transition. It is worth noting that the symmetry of the bcc (110) surface allows biaxial anisotropy, but this was not considered here in order to simplify the theoretical calculations. A thorough investigation of the same model, but with biaxial anisotropy and different directions of the DM vector is given in \cite{Heide}. It was proven that if the DM vector is parallel to the easy axis, but the plane perpendicular to this axis is not isotropic, then there is a range for the magnitude of the DM vector where the ground state ordering is neither collinear nor a planar spin spiral, but a truly three-dimensional structure, where both $m^{2}(\boldsymbol{k}=\boldsymbol{0})$ and $m^{2}(\boldsymbol{k}=\boldsymbol{k}_{min})$ are finite. The calculations in \cite{Heide} are performed only at $T=0$, but allowing biaxial anisotropy in the model may certainly influence the finite-temperature behaviour as well. However, we anticipate that this does not change the main result of the current calculation, namely the temperature-induced transition into the non-collinear state.

The transition from a non-collinear ground state to a collinear ordering may not be explained only by increasing the temperature in the current model. Including an external magnetic field $\boldsymbol{B}$ in (\ref{Ham}) parallel to the easy axis increases all magnon frequencies by the Larmor frequency, $BM$. Consequently, a suitably large value of $B$ may reorder the system, switching from the spiral state to the ferromagnetic phase. When the system is heated up at this value of $B$, our simulations indicated that again a different phase appeared at a sufficiently high temperature, where both $m^{2}(\boldsymbol{k}=\boldsymbol{0})$ and $m^{2}(\boldsymbol{k}=\boldsymbol{k}_{min})$ are finite, similarly to the three-dimensional spin structure discussed in the case of biaxial anisotropy in \cite{Heide}.

Another important observation is that the incommensurate spiral state has a continuous rotational symmetry, that is if every spin vector is rotated with the same phase in the plane, the total energy remains the same since the exchange and DM terms depend only on the relative orientation of the spins, while the anisotropy term is zero for the spiral state. At finite temperatures this leads to diverging transverse susceptibility, which prohibits any kind of long-range order in a two-dimensional system such as the one discussed in this paper, similarly to the absence of ferromagnetic or antiferromagnetic order in isotropic two-dimensional magnets at finite temperature proven by the well-known Mermin-Wagner theorem\cite{Mermin}. Including biaxial anisotropy may break this continuous symmetry, but due to the incommensurate nature of the spiral state the spiral may still relatively freely move along the lattice, leading to the disappearance of long-range order. In the finite-sized systems on which the simulations were carried out, this instability is noticeable in the behaviour of the variance of the magnetization
\begin{eqnarray}
\Delta m^{2}(\boldsymbol{k})=m^{2}(\boldsymbol{k}) - \sum_{\alpha=x,y,z}\left|\langle\frac{1}{N}\sum_{j}e^{i\boldsymbol{k}\boldsymbol{R}_{j}}\sigma_{j\alpha}\rangle\right|^{2}, \label{var}
\end{eqnarray}
which should be small in both the ordered and disordered phases, only showing a peak near the critical temperature where the order parameter of the corresponding $\boldsymbol{k}$ vector disappears. On the contrary, in the simulations $\Delta m^{2}(\boldsymbol{k}=\boldsymbol{k}_{min})$ took a large value in the spin spiral phase, indicating that the order probably disappears in an infinite system.

On the other hand, actual experiments such as SP-STM are not carried out in infinite systems, but generally on thin stripes of monolayers on the substrate, where the finite width of the stripes has an important effect on the magnetic order as shown in \cite{Meckler2} and \cite{Sessi}. In order to take these effects into account in our simulations, we used periodic boundary conditions in one direction and free boundary conditions in the other direction, with an extra on-site anisotropy term in the plane of the spiral for the atoms on the free boundary. In these simulations, the order parameter $m^{2}(\boldsymbol{k}=\boldsymbol{k}_{min})$ behaved similarly to the infinite case, but the variance $\Delta m^{2}(\boldsymbol{k}=\boldsymbol{k}_{min})$ became small in a wide temperature range, indicating that the spiral is frozen into the system at these temperatures instead of freely rotating.

\section{Conclusion}

We examined a model Hamiltonian (\ref{Ham}) describing a magnetic monolayer on a bcc (110) surface. Finite-temperature spin wave calculations indicated that the DM interaction may considerably decrease the temperature where the collinear order disappears. By performing atomistic spin dynamics simulations, it was confirmed that at this temperature the system turns into an incommensurate spin spiral state instead of the paramagnetic state. The stability of this spiral state was also analyzed, concluding that the finite size of the monolayers used in the experiments strongly influences it. Further research may turn towards a more thorough examination of this dependence of stability on the system size.

Another possibility to extend this research is using ab initio calculations, for example the relativistic torque method\cite{Udvardi2}, to determine the parameters in the model Hamiltonian. The present calculations indicated that increasing the DM interaction or decreasing the easy-axis anisotropy both bring the system closer to the spiral state. The anisotropy is strongly dependent on a number of parameters which may be changed in ab initio calculations and to some extent in experiments as well, such as the number of atomic layers, the relaxation between the magnetic layer and the substrate and the type of atoms in the substrate. For example, it is known\cite{Hauschild} that for monolayer Fe on W(110) the easy magnetization axis is in the plane of the film, while for double-layer Fe it is out of plane, probably contributing to the change in the type of ordering. Decreasing the distance between the Fe and the top W layer or replacing W with Ta may lead to a similar reorientation of the easy axis or at least weakening of the on-site anisotropy, leading to the sequence of phase transitions described in this paper.

\section*{Acknowledgment}

This work was supported by the European Union under FP7 Contract NMP3-SL-2012-281043 FEMTOSPIN. The work of LS was additionally supported by the European Union, co-financed by the European Social Fund, in the framework of T\'{A}MOP 4.2.4.A/2-11-1-2012-0001 Hungarian National Excellence Program.

\end{document}